\documentstyle[epsf]{elsart}
\def\gtrsim{ \> \lower4pt\hbox{${\buildrel > \over \sim}$} \> }
\def\lesssim{ \> \lower4pt\hbox{${\buildrel < \over \sim}$} \> }

\begin{document}
\begin{frontmatter}

\title{Variability in Blazars}

\author{Charles D. Dermer}

\address{E. O. Hulburt Center for Space Research, Code 7653 \\
Naval Research Laboratory, Washington, DC 20375-5352}

\begin{abstract}

The kinetic energy of bulk relativistic plasma ejected from the
central engine of blazars is converted into
nonthermal particle energy in the comoving frame
 through a process of sweeping up
material from the surrounding medium. The resulting deceleration of the
bulk plasma introduces a number of effects which must be included in blazar
modeling.   For example, the varying Doppler factor means that model fits
must employ time integrations appropriate to the observing times of the
detectors. We find that the ratio of the peak synchrotron fluxes reached
at  two different photon energies is largest when viewing along the jet
axis, and becomes smaller at large angles to the jet axis. This effect is
important in studies of the statistics of jet sources. Variability due
either to bulk plasma deceleration or radiative cooling must be
distinguished in order to apply recently proposed tests for beaming from
correlated X-ray and TeV observations. The blast-wave physics developed
to analyze these problems implies that most of the energy injected in the
comoving frame is originally in the form of nonthermal hadrons. Because
plasmoid deceleration can produce rapid variability due to a changing
Doppler factor, arguments against hadronic blazar models related to the
long radiative cooling time scale of hadrons are not compelling.

\end{abstract}


\end{frontmatter}

\noindent{Invited talk at VERITAS Workshop on TeV Astrophysics of 
Extragalactic Sources, Cambridge, MA, Oct. 23-24, 1998. In press,
{\it Astroparticle Physics}, ed. M. Catanese, J. Quinn, T. Weekes}

\section{Introduction}
Ejection of relativistic plasma from a compact central engine is thought to
account for the appearance and observational properties of a number of
fascinating systems in astronomy, including galactic black hole jet
sources, radio galaxies, quasars, and gamma-ray bursts (GRBs). Several
arguments have led to this conclusion, perhaps the most important being
measurements of apparent transverse superluminal motion in multi-epoch VLBI
observations of radio quasars at the sub-pc scale.  Apparent transverse
speeds exceeding $\sim 10$c are found in many sources (e.g., Vermeulen
\& Cohen 1994). The interpretation of these observations in terms of bulk
plasma outflow is not conclusive, however, as this effect could be related
to the pattern speed of the emitting regions rather than to bulk plasma
ejection.  Another argument for relativistic plasma outflow yields a
 lower limit to the Doppler factor ${\cal D}$
from the measured radio flux density, the angular diameter of the radio
emission region, and the upper limit of the self-Compton X-ray flux (e.g.,
Marscher 1987; Ghisellini 1989). An accurate measurement of
${\cal D}$ through this method requires contemporaneous radio and X-ray
measurements and, moreover, an accurate determination of the radio
self-absorption frequency.  These conditions are met only rarely, but do
point to relativistic motions in some flat-spectrum
radio quasars.

Other tests for beaming try to establish conditions for the
impossibility of intense, rapidly variable emission from stationary
radiation sources. The argument of Elliot \& Shapiro (1974) contrasts the
range of allowed black hole masses for luminosities governed by
Eddington-limited accretion, and variability time scales constrained by the
light-travel time across a region with dimensions corresponding to the
Schwarzschild radius of the black hole.  A stationary, Eddington-limited
emitting region is not possible if
$L_{48}/\Delta t({\rm days}) \gg 1$, where $10^{n}L_{n}$ ergs
s$^{-1}$ and $\Delta t({\rm days})$ are the observed luminosities and
variability time scales in days, respectively. Klein-Nishina corrections
must be applied (Dermer \& Gehrels 1995) for observations at $\gamma$-ray
energies.

Opaqueness of the emitting region to $\gamma$-$\gamma$
attenuation has also been used to argue in favor of beaming (Maraschi et
al.\ 1992).  In its simplest form, the constraint from the compactness
parameter
$\ell = 4\pi m_ec^3/\sigma_T = 6.6\times 10^{29}$ ergs s$^{-1}$ cm$^{-1}$
implies that if $L_{45}/\Delta t({\rm days})
\gg 1$, then beaming is implied.  Here the luminosity refers to emission
near 1 MeV where the pair attenuation cross section is largest;
consequently this test is most sensitive for observations near 1 MeV. 
Otherwise, assumptions about the cospatial origin of gamma rays and lower
energy radiation must be justified by observations of correlated
variability, since the cross section of
$\gtrsim 100$ MeV photons with each other is negligible (Dermer \&
Gehrels 1995).

Correlated X-ray and TeV observations of Mrk 421 (Macomb et al.\ 1995;
Buckley et al.\ 1996) and Mrk 501 (Catanese et al.\ 1997)  have demonstrated
that 2-10 keV X-rays and $\gtrsim 300$ GeV $\gamma$ rays originate from the
same region, verifying the cospatial assumption for these sources.  Their
luminosities are not large enough to establish beaming through
$\gamma$-$\gamma$ transparency arguments, but can be used to determine the mean
magnetic field $H$ in the emitting region and
establish a lower limit to the Doppler factor ${\cal D}$ through a newly
proposed beaming test. This test is described in more detail below.

Because $\gamma$-ray observations probe the region nearest the black
hole, it is important to critically examine these tests. An early hope
was that such measurements could discriminate between accelerating and
decelerating jet models (e.g., Marscher 1999) by charting the variation of
${\cal D}$, thereby revealing whether the evolution of a blazar flare is
accompanied by a prompt phase of Doppler variation.
The possibility that the Doppler factor of the emitting region
can change, however, introduces intrinsic variability which must be
distinguished from variability produced by radiative cooling of the
emitting particles. It is therefore important to consider
processes which change the bulk Lorentz factor of the radiating plasma.  This
is done in Section 2.  In Section 3, we present numerical simulation results
showing the effects of bulk plasma (or plasmoid) deceleration.  Implications
for beaming tests and blazar models are discussed in Section 4.

\section{Blast-Wave Physics}

Crucial for understanding variability behavior in blazars is
to  treat the injection of relativistic nonthermal particles in the comoving
plasma fluid frame properly. The blast-wave physics developed to model
GRB afterglows (e.g., Vietri 1997; Waxman 1997; Wijers et al.\ 1997) offers 
a solution to 
the particle injection problem, and provides a method to deal with the
deceleration of the emitting plasma. The basic idea is that the energy of the
injected nonthermal particles comes at the expense of the directed bulk
kinetic energy of the fluid. The variation of the bulk Lorentz factor
$\Gamma$ of the radiating fluid can be simply
obtained in a one-zone approximation through a momentum conservation
equation  (Dermer \& Chiang 1998). 

Suppose that the system produces an outflow with total energy $E_0$ and
initial bulk Lorentz factor $\Gamma_0$. Because most of the energy of the
flow is bound up in the kinetic energy of baryons, assumed here to be
protons, then $E_0 =
\Gamma_0 N_{\rm th} m_pc^2$, where $N_{\rm th}$ is the total number of
protons.  Thus $\Gamma_0$ represents the baryon loading of the system.

It is straightforward to write a conservation equation for the radial (or
$\hat x$) momentum component of the fluid, given by 
$$\Pi_x(x) = m_p c P \{ (1+a) N_{\rm th} + \int _0^\infty dp\cdot
\gamma\cdot [N_{\rm pr}(p;x) + a N_{\rm e}(p;x)] \}\;,\eqno(1)$$
where $a \equiv m_e/m_p$, $P \equiv B\Gamma = \sqrt{\Gamma^2-1}$, and
$p = \beta\gamma= \sqrt{\gamma^2-1}$.  The functions $N_{\rm k}(p;x)
\equiv  d N_{\rm k}(p;x)/dp$ represents the comoving distribution functions
of particles of type k = pr (protons) or k = e (electrons) at location $x$.
This expression assumes no particle escape. Eq.\ (1) indicates that the
momentum of the bulk plasma consists of both the inertia from the
thermal protons associated with the baryon loading of the explosion, and
the inertia bound up in the nonthermal proton and electron distributions. 
The latter functions evolve when nonthermal particles are injected into
the plasma and when the energy of the nonthermal particles is radiated.

As a plasmoid or blast wave traverses the surrounding medium, it intercepts
and sweeps up material. A proton and electron pair is captured by the
plasmoid with a Lorentz factor $\Gamma$ in the comoving plasma frame. The
plasmoid captures protons and electrons from the surrounding medium at the
rate
$$dN_{\rm pr,sw}(p,x)/dx = dN_{\rm e,sw}(p,x)/dx = n_{\rm ext}(x) A(x)
\delta(p-P)\;,\eqno(2)$$
The quantity $n_{\rm ext} (x)$ is the density of particles in the
surrounding medium, and $A(x)$ is the cross-sectional area of the
plasmoid which is effective at sweeping up material from the external
medium. The power of nonthermal particle kinetic energy injected into the
comoving frame is simply
$$\dot E_{\rm ke} = m_pc^2\int _0^\infty dp\;(\gamma -1)\;[\dot N_{\rm
pr,sw}(p,x) + a\dot N_{\rm e,sw}(p,x)]\;,\eqno(3)$$
where the time derivative refers to time in the comoving frame.  The
distance $\delta x$ traveled during the comoving time interval $\delta t$
is 
$\delta x = \delta t/(B\Gamma c)$.  Eqs.\ (2) and (3) therefore imply
$$\dot E_{\rm ke} = m_pc^2 B\Gamma (\Gamma - 1) (1+ a) c n_{\rm ext}(x)
A(x)\;\eqno(4)$$ 
(Blandford \& McKee 1976). It is important to note that the
fraction
$1/(1+a)=$ 99.95\% of the energy initially injected into the comoving
frame is carried by protons, though plasma processes can be
effective at transforming this energy to electrons or magnetic field.

By using eq.\ (1) to write $\Pi_x(x+\delta x)$, to which is added a
term $(d\Pi_x^{\rm rad}/dx)\delta x$ to account for radiation losses, one
obtains an equation of motion for the dynamics of the blast wave by
expanding to first order in $\delta x$ and using momentum conservation,
i.e., $\Pi_x(x+\delta x) = \Pi_x (x)$.  It is
$${-dP(x)/dx\over P(x)} = {n_{\rm ext}(x) A(x)\Gamma(x) \over 
(1+a) N_{\rm th} + \int _0^\infty dp\cdot
\gamma\cdot [N_{\rm pr}(p,x) + a N_{\rm e}(p;x)]}\;. \eqno(5)$$
If external Compton scattering processes operate, an additional term must
be added to take into account the momentum impulse from the scattered
external photons (B\"ottcher \& Dermer 1999).

Eq.\ (5) is the basic equation for calculating the dynamics of a
plasmoid by sweeping up material from the surrounding medium, and can be
solved in a number of limiting cases.  In the
relativistic  limit ($\Gamma
\gg 1$) and the blast-wave case where the area $A(x)\propto x^2$, there
are two important regimes:  the adiabatic (or non-radiative) and radiative
 regimes, where the swept-up particles retain all or none  of their
kinetic energy, respectively.  Considering the simplest case where the
density of the external medium can be parameterized by the expression
$n_{\rm ext} (x) = n_0 (x/x_d)^{-\eta}$, one finds that $\Gamma(x) \cong
\Gamma_0$ for $x\ll x_d$, and $\Gamma(x)\propto x^{-g}$ for
$x\gg x_d$, where
$g = 3-\eta$ and $g = (3-\eta)/2$ in the adiabatic and radiative regimes,
respectively.  The deceleration radius
$$x_d = [{(3-\eta)E_0\over 4\pi f_b n_0 \Gamma_0^2
m_pc^2}]^{1/3}\;\eqno(6)$$
(e.g., Rees \& M\'esz\'aros 1992), and represents the characteristic
distance beyond which the behavior changes from a coasting solution to a 
decelerating solution. The term $f_b$ represents the
fraction of the full sky into which the explosion energy is ejected.

For intermediate radiative regimes where a fraction $\zeta$ of the
swept-up energy is retained in the blast wave, so that a fraction
$(1-\zeta)$ is dissipated as radiation or lost through escape, Dermer et
al.\ (1999) show that $g = (3-\eta)/(1+\zeta)$. In general, one may
describe the dynamics of a blast-wave which decelerates by sweeping up
material from a surrounding medium which is distributed according to the
relation $n_{\rm ext}(x)\propto x^{-\eta}$ by an
expression of the form
$$\Gamma(x) \cong \Gamma_0/[1+(x/x_d)^g]\;.\eqno(7)$$
The relationship between the location $x$ of the blast wave and the
observing time $t_{\rm obs}$ can be obtained by noting that the radiating
element travels a distance $\delta x = {\cal D} \Gamma B c \delta t_{\rm
obs}/(1+z)$ during the observing time interval $\delta t_{\rm obs}$,
where $z$ is the redshift of the source,  ${\cal D} =
[\Gamma(1-B\cos\theta )]^{-1}$ is the Doppler factor, and $\theta$ is the
angle between the direction of motion of the radiating element (or the jet
axis) and the observer's direction. From this relation, one can show that
$x
\propto  t_{\rm obs}$ when $x \ll x_d$, and $ x \propto t_{\rm
obs}^{1/(2g+1)}$ when $x \gg x_d$.

The above expressions are sufficient to treat analytically the basic
effects from blast wave deceleration and energization through the sweep-up
process.  Assuming that the fraction $(1-\zeta)$ of the swept-up power is
dissipated in the form of radiation, then the radiated power in the
comoving frame at $x$ is $\dot E
\propto (1-\zeta)\Gamma^2(x) n_{\rm ext}(x) A(x)$.  The received power from a
portion of the blast wave directed along the line-of-sight to the
observer is equal to $\dot E$ amplified by a factor $\Gamma^2(x)/(1+z)^2$ due
to the transformations of energy and time. If the observer is outside the
Doppler cone of the plasmoid, the emission is weak until the radiating plasma
has slowed down sufficiently so that the Doppler cone intercepts the line of
sight. At later times, the emission approaches the behavior found in the
case where the observer's line-of-sight intercepts the radiating region.

If $\psi$ denotes the opening angle of the plasmoid (or jet), 
then two limits are
important when the area of the plasmoid increases
$\propto x^2$.   From the above discussion, the received
bolometric power
$P(t_{\rm obs})
\propto (1-\zeta)\Gamma^4(x) n_{\rm ext}(x) A(x)/(1+z)^2$. The
dynamics of the blast wave changes from a coasting solution to a
decelerating solution when it passes the deceleration radius $x_d$, which
occurs at the observing time
$$t_d = \Gamma_0(1-B_0\cos\theta)(1+z) x_d/(c\Gamma_0) \rightarrow
(1+z)x_d/(2\Gamma_0^2 c)\;;\eqno(8)$$
here $B_0 = \sqrt{1-\Gamma_0^{-2}}$ and the right-hand expression of eq.\ (8)
refers to the case when the plasmoid is directed along the line-of-sight.

For the case $\theta \lesssim \psi$, $P_p(t_{\rm obs})\propto t_{\rm
obs}^{2-\eta}$ when  $t_{\rm obs} \ll t_d$,  and $P_p(t_{\rm obs})\propto
t_{\rm obs}^{(2-\eta-4g)/(2g+1)}$ when $t_{\rm obs} \gg t_d$. For
observations at $\theta \gtrsim \psi$, 
the emission from the blast wave begins to
intercept the observer's line-of-sight when
$\theta = 1/\Gamma(x)$, which occurs when $t_{\rm obs}
\approx t_d (\Gamma_0\theta)^{(2g+1)/g}$. At this time, the received power
is nearly equal to the value supposing that $\theta \lesssim \psi$.

We therefore see that the simplest model employing blast-wave physics
shows how a plasmoid is energized by
sweeping up material and converting it into nonthermal particle energy in
the comoving frame (see Dermer et al.\ 1999
for more details). For a jet directed along the observer's line-of-sight
(which is generally thought of as the standard model for blazars), the
sweeping-up process produces a flare with bolometric flux rising
$\propto t_{\rm obs}^{2-\eta}$.  After a sufficient amount of
material has been swept up to cause the blast wave or plasmoid to
decelerate, the Doppler deboosting overpowers the additional
energization to cause the received flux to decay $\propto t_{\rm
obs}^{(2-\eta-4g)/(2g+1)}$.  For adiabatic ($g = 3/2$) and radiative ($g =
3$) blast waves in a uniform surrounding medium with $\eta = 0$, the light
curves decay $\propto t_{\rm obs}^{-1}$ and $\propto t_{\rm
obs}^{-10/7}$, respectively. The decaying flux is entirely a consequence
of the decreasing Doppler factor. Thus it is not valid to interpret a
decaying flux as evidence for cooling of the emitting particles without a
discriminant between the effects of cooling and deceleration.

\section{Numerical Calculations}

\setcounter{figure}{0}
\begin{figure}
\vskip -7 truecm
\epsfysize=14cm
\epsffile[60 180 600 700]{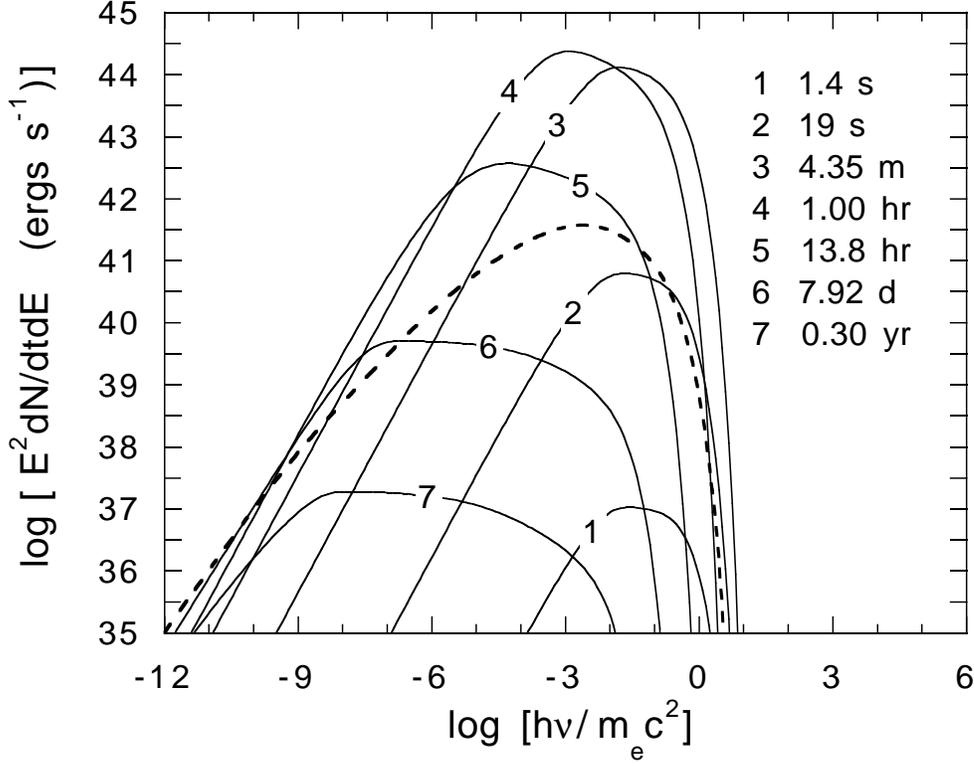}
\vskip 3.0 truecm
\caption[]{Calculations of synchrotron spectra at times given
by the labels on the figures. Observer is viewing down the jet axis.
Parameters of the calculation are given
in the text.  Dashed curve shows the spectrum averaged over a one-day
integration time.
 }
\end{figure}

Using the code described in Chiang \& Dermer (1999), we can illustrate
the effects described in the previous section for a pure synchrotron
flare.  In the simulation shown in Fig.\ 1, we
 let the central engine eject $10^{48}$ ergs of plasma with a baryon loading
given by $\Gamma_0 = 40$ into 10\%  of the full sky. Thus the opening angle
of the jet is
$11.5^\circ$. The jet plasma passes through a medium with a uniform density
of  0.01 cm$^{-3}$, and as it sweeps up this material it converts it 
with high efficiency into
nonthermal power-law electrons with a number injection index 
$dN/d\gamma \propto \gamma^{-3}$. The energy of the injected
 electrons ranges between $\gamma = \Gamma$ and 1\% of the maximum energy given by 
balancing the electron synchrotron loss time scale
 and the Larmor time scale in a magnetic field. We assume that the magnetic field
energy density is 10\% of the
 downstream energy density of the swept-up
particles. This represents a magnetic field of 0.5 Gauss during the phase
before the blast wave begins strongly to decelerate. 

Electron synchrotron cooling is taken into account in the calculation, but
makes only a small contribution to the variability shown here, 
which is due overwhelmingly
to energization of the plasmoid by sweeping up particles,
 and to the subsequent deceleration
and Doppler deboosting that results from this process. 
The general progress of the flare for the given parameters is to rise
rapidly at X-ray and soft $\gamma$-ray energies.  The flare then sweeps to
lower energies on a much longer time scale. Note that the flare reaches
larger $\nu F_\nu$ peak fluxes at higher energies where it is most variable. 
At lower photon energies, the variability is less extreme, and the 
peak $\nu F_\nu$ value reached is lower.  

The dashed curve in Fig.\ 1
shows a one-day time integration over the flaring emission.
As can be seen, the time-integrated spectrum is 
much softer than the time-resolved
spectra because it represents a superposition of hard spectra which peak at
successively lower energies. When fitting blazar flare data using a model
such as described here, it is therefore necessary to perform 
time-integrations over
the flaring emission appropriate to the sampling time of the detector. 
Because gamma-ray telescopes require long observing periods to 
accumulate sufficient statistics,
this is especially important for jointly fitting hard X-ray/soft
$\gamma$-ray data, or MeV-GeV flares resulting from the synchrotron
self-Compton (SSC) or external Compton scattering process.

\setcounter{figure}{1}
\begin{figure}
\vskip -7 truecm
\epsfysize=14cm
\epsffile[60 180 600 700]{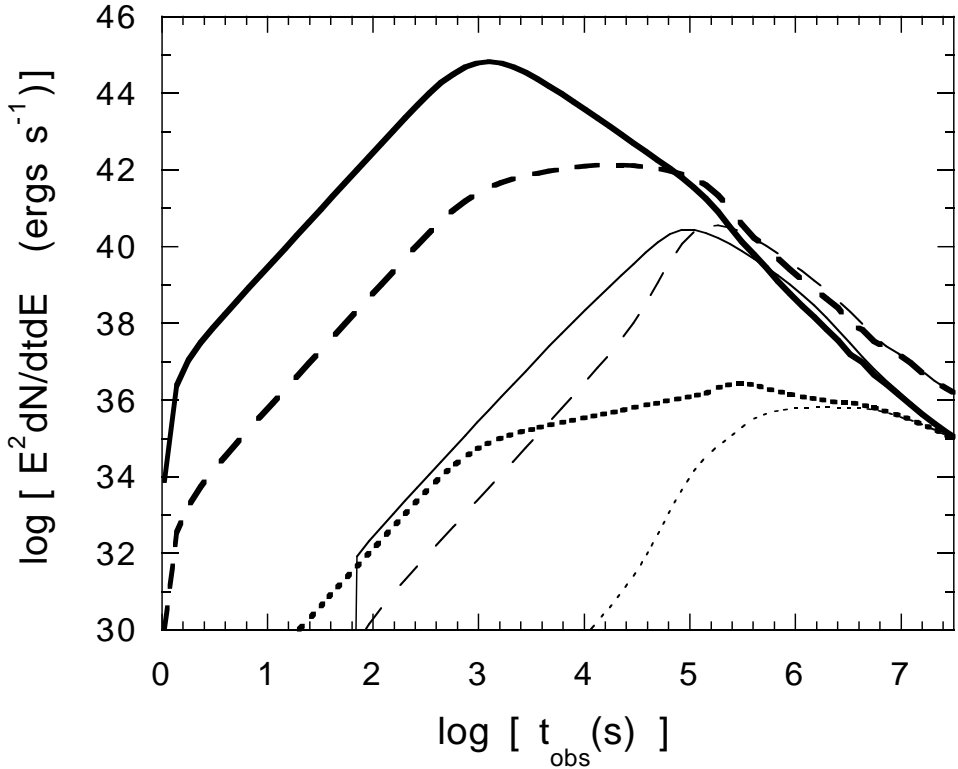}
\vskip 4.0 truecm
\caption[]{ Calculations of synchrotron flare light curves using the same
parameters as in Fig. 1. Solid, dashed, and 
dotted curves are 1 keV X-ray, 1 eV optical, and 2.4 GHz radio light curves,
respectively. Thick curves correspond to observations down the axis of the jet
with  opening angle $\psi =
11.5^\circ$, and thin curves represent observations at $\theta = 20^\circ$ 
from the jet axis.
 }
\end{figure}

\setcounter{figure}{2}
\begin{figure}
\vskip -7 truecm
\epsfysize=14cm
\epsffile[60 180 600 700]{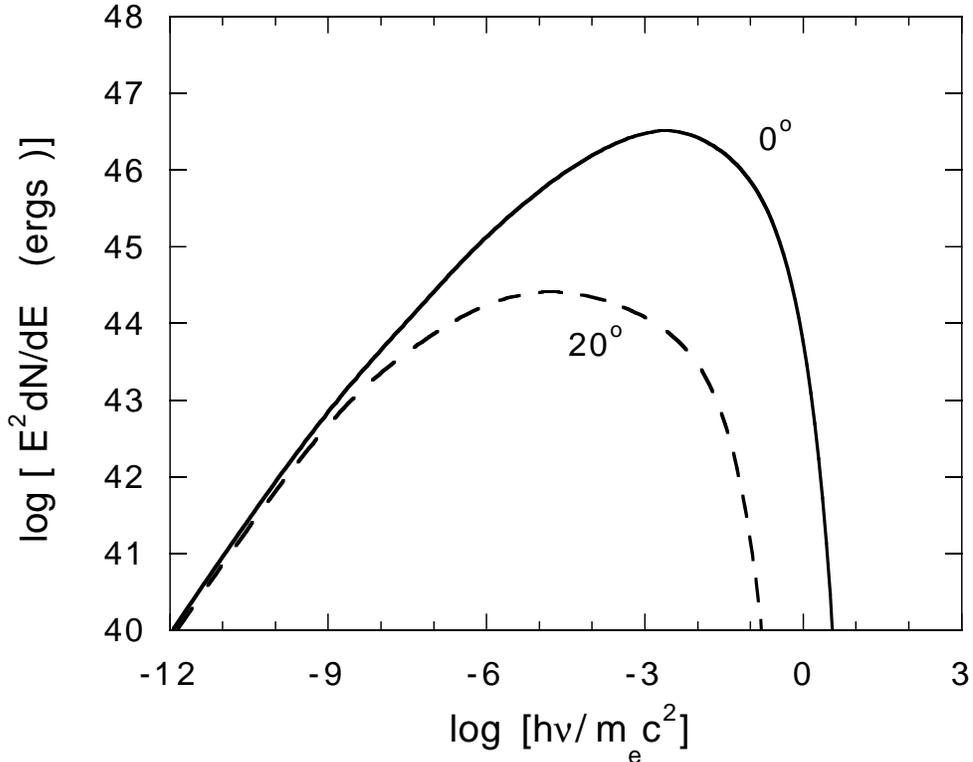}
\vskip 3.0 truecm
\caption[] { Total energy received at different observing energies
for the flare parameters shown in Fig.\ 1, for observations along the
jet axis (solid curve) and at 20$^\circ$ off the jet axis (dashed
curve).  Flux-limited telescopes will preferentially detect jet
sources along the jet axis when observing at the highest photon energies.
 }

\end{figure}
Fig.\ 2 shows light curves for the model synchrotron flare shown in
Fig.\ 1 at X-ray, optical and radio frequencies, both along the jet axis
($\theta = 0^\circ$; thick curves),
and at $20^\circ$ to the jet axis (thin curves).
 When observing along the jet axis, the 
peak $\nu F_\nu$ flux measured at higher frequencies 
is much greater than the peak $\nu F_\nu$ flux measured 
at lower frequencies. By contrast, when observing outside the 
opening angle of the jet, i.e., when $\theta > \psi$, one sees that the range 
of peak fluxes becomes much less.  Consequently, a flux-limited telescope
 observing at higher photon energies will be much more likely to 
detect beamed sources along the jet axis than a flux-limited telescope 
sensitive to lower energies, which will detect on-axis and off-axis sources
with nearly equal likelihood.

This effect is a consequence of the energization and deceleration of the 
radiating plasma which causes the received emission to sweep to lower energies
at late times, and has important implications
 for any statistical analyses of jet sources. In usual statistical treatments, 
one generally assumes that the relative flux observed at
different angles to the jet axis is governed by the factor 
${\cal D}^{3+\alpha}$, {\it independent of photon energy} 
(here $\alpha$ is the 
energy index  of the flux density $F_\nu \propto \nu^{-\alpha}$; see, e.g.,
Urry \& Padovani 1995). As shown here, the situation is much more complicated
for flaring sources, which blazars most certainly are.  A flux-limited X-ray
survey 
 primarily detects X-ray jet emission from aligned sources,
with off-axis jet sources being too faint to detect at X-ray energies. 
 By contrast, a flux-limited survey at radio energies will detect
on-axis and off-axis sources at comparable flux levels. Thus we expect and do
see the parent population of radio quasars, namely the radio galaxies.  The
parent population of X-ray selected BL Lac objects, by contrast, are at such 
a low flux level to hardly be detectable.

Fig.\ 3 shows the integrated energy fluence measured from a blazar flare at
different photon energies, illustrating the effect just described.
We have considered only the synchrotron emission up to this point, but the 
same behavior operates for the SSC flux 
(Dermer, Mitman, \& Chiang 1999, in preparation).
 Variability will be greatest at higher photon energies,
where the largest $\nu F_\nu$ fluxes are reached for on-axis sources.
The variability time scale will be longer and the flux roughly equal
for aligned and misdirected blazars at lower photon energies. This same
general behavior probably applies to external Compton scattering emission
as well,
though additional complications from the narrower beaming cone of ECS
compared to the synchrotron and the SSC processes (Dermer 1995)
must be taken into account in this case.

\section{Beaming Tests and Blazar Models}

As noted in the Introduction, the correlated X-ray and TeV observations
provide a new test for beaming in blazars. This test has been recently
applied to observations of Mrk 501 by Catanese et al.\ (1997), Dermer
(1998), and Kataoka et al.\ (1999).  In this test, it is assumed that the X-ray
emission is nonthermal synchrotron emission,  and that the variability is a
result of synchrotron losses in a magnetic field of mean intensity $H$.  If
the minimum variability time scale measured at energy $\bar E$ is denoted by
$\delta t_{\rm obs}^{\rm min}$, then 
$$H({\rm G}) \cong 0.8\;\{ {(1+z)\over {\cal D} \bar E({\rm keV}) 
[\delta t_{\rm obs}^{\rm min}({\rm
hr})]^2 }\}^{1/3}\;.\eqno(9)$$ 
(e.g., Tashiro et al.\ 1995; Takahashi et al.\ 1996).

An upper limit on the mean magnetic field
$H$ in the emitting region  is implied because the electrons producing the
highest energy synchrotron emission have Lorentz factors
$\gtrsim 2\times 10^6 E_{\rm C}({\rm TeV})
 (1+z)/{\cal D}$, where $E_{\rm C}({\rm TeV})$ is the
measured energy in TeV of the highest-energy gamma-rays. Synchrotron emission 
correlated with the TeV flux requires that the electrons radiate in a magnetic field
at least as great as 
$H({\rm G}) \cong 11 \epsilon_{\rm obs, syn}{\cal D}/[ E^2_{\rm C}
({\rm
TeV}) (1+z)]$, where $\epsilon_{\rm obs, syn}$ is the measured dimensionless energy of the
highest energy synchrotron photons produced by the electrons which produce the
TeV radiation. When compared with the value of
$H$ inferred through equations (9), one obtains an
expression for the Doppler factor, given by 
$${\cal D}
 \cong 1.7\;{(1+z)  [E_{\rm C}({\rm TeV})]^{3/2}\over \bar \epsilon_{\rm
obs}^{1/4}({\rm keV}) (\delta t_{\rm obs}^{\rm min})^{1/2}(\epsilon_{\rm obs,
syn})^{3/4}}\;\eqno(10)$$
(Dermer 1998). A lower limit to ${\cal D}$ is obtained if the TeV
flux does not exhibit a clear cutoff due to the high-energy cutoff in the
electron distribution function.

With the advent of air Cherenkov telescopes
such as Whipple and HEGRA detecting BL Lac objects to TeV and tens of TeV
energies, this test could in principle provide the largest inferred Doppler
factors of all known beaming tests (J. H. Buckley, 1998, 
private communication).  It is necessary, however, 
to discriminate clearly between bulk
deceleration effects and radiative cooling effects.  Unfortunately, the
effects of deceleration mimic those of radiative cooling in a variety of ways
(Chiang 1999). This is true both for the energy-dependent time lags produced
by synchrotron cooling, and for the clockwise loop diagrams produced by
flaring sources when data is plotted in a spectral index/intensity
display (see also Kirk et al.\ 1999). One such discriminant might be the
slower decay of the SSC emission compared to synchrotron emission (Dermer
1998), but a detailed numerical simulation will be required to fully resolve
this question.

Finally, we note that the existence of the process of plasmoid deceleration
weakens arguments (Buckley 1998) against hadronic models based on the long
cooling time scales of protons through photo-meson, photon-pion, and proton
synchrotron processes (e.g., Biermann \& Strittmatter 1987) compared to
the observed rapid variability time scales of BL Lacs.  As shown here,
neither extremely high energy protons nor intense radiation fields are
required to produce rapid variability, which can result solely from
Doppler deboosting due to the deceleration of the radiating region. This, and
the fact that most of the nonthermal particle energy injected into the
plasmoid is initially in the form of relativistic protons according to the
blast wave physics described here, seems to give new life to hadronic models
of blazars (e.g., Mannheim 1993; Gaisser et al.\ 1995) 
provided, of course, that hadronic models can
successfully fit the generic two component
$\nu F_\nu$ blazar spectrum.  If radiative cooling
does not produce the variability behavior, hadronic models must also
 contend with low radiative efficiencies in an uncooled model (see B\"ottcher \& Dermer 1998; Totani 1998 for related questions in GRB models).
 The variability behavior normally attributed to
radiative cooling processes would, in the hadronic models, instead be a
consequence of plasmoid deceleration.
 Thus a discriminant between cooling and
deceleration effects is also of central importance
 to distinguish leptonic and
hadronic models of blazars. 

\section {Summary}

Recent observations of power-law X-ray afterglow observations of GRBs
 have stimulated the development of new physics
for calculating radiation from relativistic plasma outflows. The
 energization of nonthermal particles in the radiating
plasma occurs through a process of
sweeping up material from the surrounding medium. The internal nonthermal
particle energy is extracted
 from the directed kinetic energy of the bulk plasma, causing the bulk
plasma to decelerate.  We have examined idealized flaring behaviors produced
 when bulk plasma sweeps  up particles from a uniform medium. 
Inhomogeneities in the surrounding medium as well as in the relativistic
plasmoid will complicate the situation.

We have shown that 
\begin{itemize}
\item Bulk plasma deceleration effects must
be included in models and statistical studies of blazars;
\item
Time integrations over the varying Doppler factor
of the radiating plasma must be performed when
fitting blazar data, employing integration ranges appropriate to the
observing times of the different telescopes;
\item A newly developed test for beaming in blazars using correlated X-ray and
TeV variability observations must answer the criticism that the variability
is due not to radiative cooling but rather to Doppler deboosting;
\item Hadronic models for blazars do not have to confront the difficulty of
the long radiative cooling time scales of hadrons, since variability can be
achieved through plasmoid deceleration.
\end{itemize}
Important for further progress on these questions is an observational
discriminant between cooling and deceleration processes.

\vskip 1 true cm
{\bf Acknowledgments}

I acknowledge valuable discussions on  blast-wave physics and synchrotron
radiation with Jim Chiang,  Markus B\"ottcher, and
Hui Li, and thank Jim Chiang for the use of his code.
This work was supported by the Office of Naval Research and the Compton
Observatory Guest Investigator Program

\end{document}